# Modeling ant battles by means of a diffusion-limited Gillespie algorithm

**Short title: Diffusion-limited ant battles model**

Gianluca Martelloni[1,4] [*], Alisa Santarlasci[2,4], Franco Bagnoli[1,4,5], Giacomo Santini[3]

[1] Department of Physics and Astronomy, University of Florence, via G. Sansone, 1 50019 Sesto Fiorentino, Firenze (Italy).

[2] Department of Information Engineering, University of Florence, via S. Marta, 1 50139 Firenze, Italy.

[3] Department of Biology, University of Florence, via Madonna del Piano, 6 50019 Sesto Fiorentino, Firenze (Italy).

[4] Center for the Study of Complex Dynamics, University of Florence.

[5] INFN, Florence

**Abstract**

We propose two modeling approaches to describe the dynamics of ant battles, starting from laboratory experiments on the behavior of two ant species, the invasive *Lasius neglectus* and the authocthonus *Lasius paralienus*. This work is mainly motivated by the need to have realistic models to predict the interaction dynamics of invasive species. The two considered species exhibit different fighting strategies. In order to describe the observed battle dynamics, we start by building a chemical model considering the ants and the fighting groups (for instance two ants of a species and one of the other one) as a chemical species. From the chemical equations we deduce a system of differential equations, whose parameters are estimated by minimizing the difference between the experimental data and the model output. We model the fluctuations observed in the experiments by means of a standard Gillespie algorithm. In order to better reproduce the observed behavior, we adopt a spatial agent-based model, in which ants not engaged in fighting groups move randomly (diffusion) among compartments, and the Gillespie algorithm is used to model the reactions inside a compartment.

## 1. Introduction

Eusociality is a characteristic of some insect species, such as bees and ants, and is the basis of their ecological success. Ants, in particular, with their worldwide distribution, are among the most abundant and ecologically relevant terrestrial group: they contribute up to 15–20% of the total terrestrial animal biomass and, due to the ability to modify their environment, are considered "ecosystem engineer" [1]. This point becomes particularly important in the case of invasive species, whose control is difficult and their eradication often practically impossible. Invasive ants are able to form vast colonies, which in some cases may merge into a single large "supercolony" of interconnected nests [2,3]. Among the key features characterizing the ecology and behavior of invasive ants, an important role is played by their ability to adopt mass fighting strategies that allow easily defeating their opponents. A deeper comprehension of the fighting behavior of invasive ant is thus needed to formulate reliable mathematical models to predict the dynamics of invasive ants spread in an ecological system.

---

[*] email: gianluca.martelloni@unifi.it

The features of invasive ants are deeply study by many points of view. Many studies take into consideration the typical aspects of invasive ants: the spatial distribution of the interconnected nests with the genetic and chemical relationships among ants [4]. The development of a spatial colony structure related to the invasive capacity of the ant species is schematized by Cremer *et al.,* [2] . Furthermore, they report a correlation analysis between the geographical distance and the dissimilarities in cuticular hydrocarbons for each pair of nests considered. Killion and Grant [5] proposed a colony-growth model for the invasive fire ant (*Solenipsis invicta*) to test the limit of its expansion range in the central United States. Fitzgerald *et al.,* [6] developed a grid-based model to predict the course of a fine-scale invasion of the invasive Argentine ants (*Linepithema humile*) from the suburban matrix to the country side. This model is based on observations of the presence/absence of ants in a given location and incorporates the invasion process from neighboring areas.

During group combats social animals may adopt different fighting strategies. In the simplest case, a battle is a sequence of duels: each individual of a group face a single opponent of the other group. In more organized armies, however, the individuals of a species can cooperate to attack a single opponent. In the latter case, group size may have a disproportionate importance over the individual fighting ability.

To describe the battles between two opposing groups, Lanchester (1916) [7] proposed two simple mathematical models, known as the "linear" and the "square" law, respectively. Under the linear law it is assumed that the battle consists of a series of duels. On the contrary, the square law considers that members of the more numerous group gang together to attack a single opponent of the less numerous group. According to the linear law, the death rates of the two species depend mainly on the fighting ability of the opponent's abundance. Under the "square law", on the contrary, the death rate of one species depends more on opponent's abundance. The Lanchester model was originally developed to model military combats, and was first applied to describe animal contests by Franks and Patridge [8]. See also [9-14].

The original Lanchester equations do not consider the spatial dimension of the battlefield. Protopopescu *et al.,*[15] included for the first time one–dimensional spatial effects in the Lanchester equations. This approach was followed by Cosner *et al.,* [16] who used a parabolic system with nonlinear interactions. Spradlin and Spradlin [17] and Keane [18,19] applied Protopopescu *et al.,* approach to two dimensions [15]. In all these papers the conflicts are modeled by means of reaction-diffusion partial differential equations. A general formulation for a spatial Lanchester model is reported by Gonzalez and Villena [20]. In their work, they pay attention to the explicit balance of forces taking into account the movements and the concentrations of opponents. None of these papers, however, take into account the behavior of single individuals during the fight and their movements in the battle area.

In order to overcome these limitations, we developed a spatial agent-based model calibrated on experimental data. The agent-based approach directly considers the interactions among the individuals composing the system. The power of this method relies on its capacity to recreate, starting by simple behavioral rules, the spatio-temporal evolution of a complex system.

For example, Luo and Opaluch [21] use a spatial-explicit agent-based simulation model to study the environmental risks of introducing the non-native oyster *Crassostrea ariakensis* in Chesapeake Bay, USA, where the native oyster populations are declining. Agent-based models are used to investigate ecological problems, where the spatial component may be important, such as spreading of invasive species (plants, animals, bacteria, virus disease etc.), dynamics of native species with references to spatial limitations, competition. See for example [22-30].

Our study focus on two species that share the same habitat but with different behavior and fighting strategies: the invasive ant *Lasius neglectus* and the autochthonous *Lasius paralienus*, both present in Europe. Asia Minor is the likely native environment of *Lasius neglectus* [31] where it derived from *Lasius turcicus*. This species, first identified near Budapest in 1990 [32] spread all over Europe and Asia [4] mainly colonizing urban habitats where it replaces most of native ant species and other arthropods [2]. As other invasive ants, *L. neglectus* forms large networks of interconnected nests, characterized by the absence of intraspecific aggression. Each nest hosts many queens (polygyny), which mate inside the nest and disperse on foot accompanied by workers (colony budding). The numerical advantage resulting from these large "supercolonies" promotes efficiency and fighting superiority with respect to other species [2, 3, 33]. *Lasius paralienus* [35], a species common in Europe, is usually more abundant in the countryside, although can also be found in urban gardens. It is a monogynous species and forms middle-sized colonies, with a limited competitive ability.

Our research started with experimental observations on the fighting behavior of these two species, staging battles where 10 *L. neglectus* confronted to 10 *L. paralienus*. Larger battles are possible but were avoided due the difficulty in following the behavior of individual ants in denser groups. We first describe the dynamics of the system by means of chemical equations which encode the interactions among individuals, *i.e.*, we consider the single ant species and the groups, formed during the battle, as chemical species. For example, two ants can attack (simultaneously or sequentially) an opponent and, if the group is sufficiently long-lasting, we consider the three-ant group a specific chemical entity. From the chemical reactions we derive a system of differential non-linear equations (DE) that gives a mean-field description of the system. The chemical rate equations are obtained by comparison, through an optimization algorithm (Flexible Simplex, [36] ), with experimental data. This deterministic approach is valid when the state variables correspond to a large number of individuals of each species [37], while in our tests the number of ants is small and indeed generate large fluctuations in the system. To take into account the finiteness of the system we use a Gillespie algorithm [38], which is an event-driven agent-based model without spatial structure. Although this approach satisfactorily reproduces the observed fluctuations in the number of groups, it may not be able to take into account of the influences of space on the dynamic of the studied system. To analyze the effect of the mobility on ant behavior and to test if the standard Gillespie algorithm is sufficient to fully characterize the experiments, we developed a spatial version of the Gillespie algorithm. We assume that ants move randomly in a disk that reproduces the geometry of our battle arenas (Petri dishes). The disk is partitioned into compartments, and the Gillespie algorithm is employed inside each compartment for choosing the next event. In this approach the reactions are limited by the diffusion, and we denote it *diffusion-limited Gillespie algorithm*.

In Section 2.1 we present the materials and the method used in experimental observations and a description of the fighting abilities of the two species considered. Section 2.2 illustrates the chemical approach, describes the deterministic model based on differential equations and explains the parameters' identification technique. In section 2.3 we introduce a stochastic model based on the classical Gillespie algorithm. In section 2.4 we implement a spatial agent-based model to test the influence of the space on the outcome of the fighting dynamics.

## 2. Materials and methods

2.1 Experimental model: Ant sampling and site

To carry out the experiments we collected ants during July/August 2012 in Prato (Northern Tuscany, Italy, 43°52' 46"N, 11°05'50"E), where colonies of *L. neglectus* and *L. paralienus* can be found in urban gardens. Each species was collected from nests at 200 m from each other. The two species are monomorphic with low intraspecific differences in ant size. After collection, specimens were maintained in a test tube, with water available, to acclimatize to laboratory conditions for one hour (temperature T ≈27 °C). Ten specimens of the two species were then simultaneously dropped into a neutral arena, represented by a 10 cm (diameter) Petri dish with Fluon © coated walls.

The interactions among ants were recorded using a fixed digital camera for five 15–minutes periods, interspaced by periods of 30 minutes during which water was offered to ants. The total recording time is therefore 75 minutes, and the total elapsed time is 3 hours and 15 minutes. We choose this procedure to obtain sufficiently long observation times, and hence increasing the probability of observing casualties, while keeping the total duration of recordings small.

The two species show different behaviors. *L. neglectus* is always the first aggressor, exhibits cooperative behavior but it also suffers greater mortality. *L. paralienus* is less aggressive, often avoids fighting, individuals do not cooperate but they are stronger than *L. neglectus* in individual duels. *L. neglectus*, on the contrary, is smaller than its counterpart and its success mainly depends on the ability of individuals to cooperate in aggressing single opponents. Groups of up to four *L. neglectus* may in fact form during the fights.

The videos were examined taking note, during time, of the number of individuals of the two species involved in a fighting group. Only specimens, remaining in close contact for at least 15 seconds, were considered as forming a group. We denote *L. paralienus* by *A* and *L. neglectus* by *B*. The groups *AB*, *ABB* and *ABBB* denote the groups in which one *A* fights with one *B*, one *A* with two *B's* and one *A* with three *B's,* respectively. During the experiments we observed that the *A* and *B* individuals, not involved in the fights, remain in motion, while ants engaged in a fight (*i.e.*, those forming a group *AB*, *ABB* or *ABBB*) do not move. We also annotated the changes in the composition of groups due to disengagement or death (a rare event) and the duration of each group. In Figure 1A-F we report the change in the number of A and B individuals and *AB*, *ABB* and *ABBB* groups observed in two experiments. Figure 1G shows the time at which we observe a new event (*i.e.*, the death of an ant or formation/ dissolution of a group). In these two experiments we observed the death of only an A individual, while for B species we observed 7 and 4 deaths.

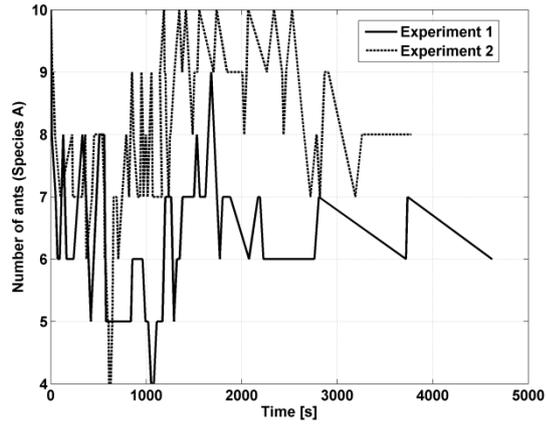
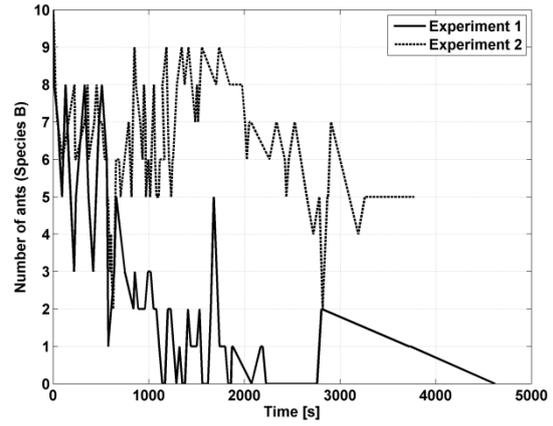
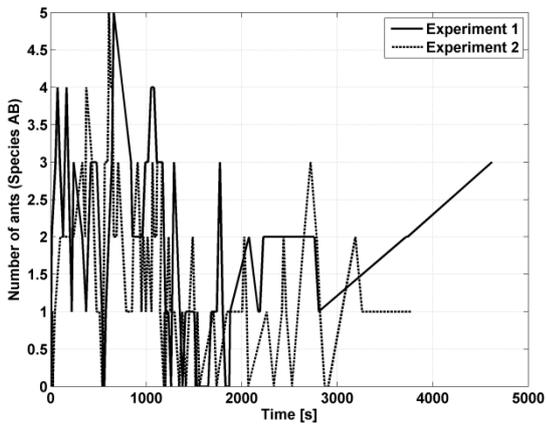
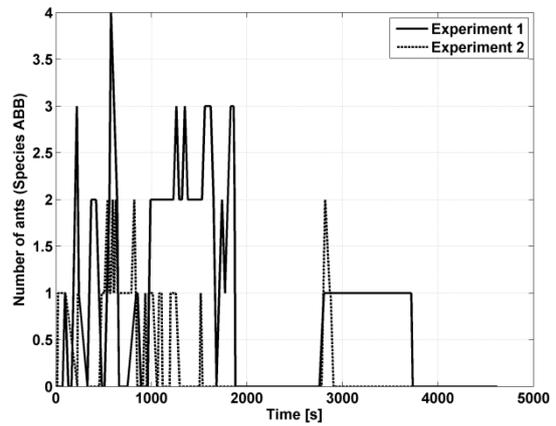
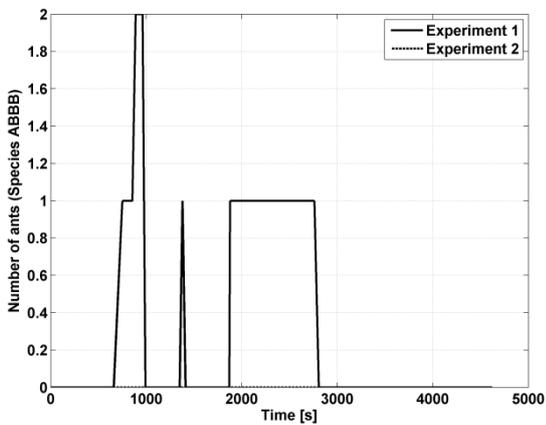
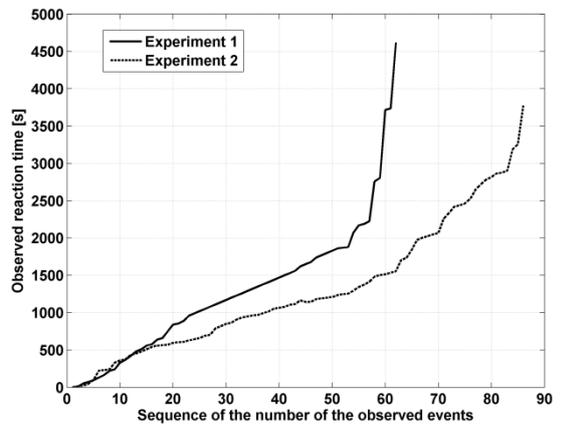

Fig. 1. The number of individuals and observed fighting groups in two experiments. (A) Number of individuals of species *A*. (B) Number of individuals of species *B*. (C) Number of AB groups. (D) Number of ABB groups. (E) Number of ABBB groups. (F) The time at which we observe a new event, *i.e.*, a death of an ant or the formation or dissolution of a group.

## 2.2 The differential model and parameter identification technique

The dynamics of the ant fighting are simulated using a chemical model which explicitly incorporates the interactions among individuals. We consider the following observed reactions:

$$A + B \underset{k_2}{\overset{k_1}{\leftrightarrow}} AB, \tag{1}$$

$$AB \xrightarrow{k_3} B, \tag{2}$$

$$AB \xrightarrow{k_4} A, \tag{3}$$

$$AB + B \underset{k_6}{\overset{k_5}{\leftrightarrow}} ABB, \tag{4}$$

$$ABB \xrightarrow{k_7} AB, \tag{5}$$

$$ABB \xrightarrow{k_8} 2B, \tag{6}$$

$$ABB \underset{k_{10}}{\overset{k_9}{\leftrightarrow}} A + 2B, \tag{7}$$

$$ABB + B \underset{k_{12}}{\overset{k_{11}}{\leftrightarrow}} ABBB, \tag{8}$$

$$ABBB \xrightarrow{k_{13}} 3B, \tag{9}$$

$$ABBB \underset{k_{15}}{\overset{k_{14}}{\leftrightarrow}} AB + 2B, \tag{10}$$

where $k_i$, $i=1,2,\ldots,15$, are the reaction constants. Equation (1) represents the formation and the dissolution of an AB group. Equations (2) and (3) describe the mortality of *A* and *B* inside an AB group. Equation (4) takes into account the formation and dissolution of group *ABB*, while Eqs. (5) and (6) model the death of *B* and *A* in larger groups. Eq. (7) describes the formation (dissolution) of an *ABB* group when two *Bs* attack (or abandon) a single *A*. Multiple attacks such as that in Eq. (7) do not occur simultaneously in real combats. However, since they may occur within a very short interval, they were considered as simultaneous. Finally Eqs. (8), (9) and (10) describe the dynamics of the chemical species *ABBB* (for more detailed see [39]).

Let $x_1$, $x_2$, .., $x_5$ denote the abundance of the chemical species *A*, *B*, *AB*, *ABB* and *ABBB*, respectively. The average dynamics of the system is described by the following system of non-linear differential equations:

$$\dot{x}_1 = -k_1 \cdot x_1 \cdot x_2 - k_{10} \cdot x_1 \cdot x_2^2 + (k_2 + k_4) \cdot x_3 + k_9 \cdot x_4, \tag{11}$$

$$\dot{x}_2 = -k_1 \cdot x_1 \cdot x_2 - 2 \cdot k_{10} \cdot x_1 \cdot x_2^2 + (k_2 + k_3) \cdot x_3 - k_5 \cdot x_2 \cdot x_3 - 2 \cdot k_{15} \cdot x_2^2 \cdot x_3 + \\ + (k_6 + 2 \cdot k_8 + 2 \cdot k_9) \cdot x_4 - k_{11} \cdot x_2 \cdot x_4 + (k_{12} + 3 \cdot k_{13} + 2 \cdot k_{14}) \cdot x_5 \tag{12}$$

$$\dot{x}_3 = k_1 \cdot x_1 \cdot x_2 - k_5 \cdot x_2 \cdot x_3 - (k_2 + k_3 + k_4) \cdot x_3 + (k_6 + k_7) \cdot x_4 - k_{15} \cdot x_2^2 \cdot x_3 + k_{14} \cdot x_5, \tag{13}$$

$$\dot{x}_4 = k_{10} \cdot x_1 \cdot x_2^2 + k_5 \cdot x_2 \cdot x_3 - k_{11} \cdot x_2 \cdot x_4 - (k_6 + k_7 + k_8 + k_9) \cdot x_4 + k_{12} \cdot x_5, \tag{14}$$

$$\dot{x}_5 = k_{11} \cdot x_2 \cdot x_4 - (k_{12} + k_{13} + k_{14}) \cdot x_5 + k_{15} \cdot x_2^2 \cdot x_3, \qquad (15)$$

where,

$$\dot{x}_i = \frac{dx_i}{dt}$$

represent the time derivatives, *i.e.*, the growth rate for each considered species.

The differential system (11-15) is integrated by means of the numerical method of Cash and Karp [40] checking that the fourth and fifth order solutions provide the same results.

The Eqs. (11-15) do not exhibit fixed points as the stationary solution ($dx_1/dt = 0$, $dx_2/dt = 0$, $dx_3/dt = 0$, $dx_4/dt = 0$, $dx_5/dt = 0$) depends on the initial conditions (Figure 2). This dependence is consistent with the biological point of view: the number of survivors of the winning species depends on the initial number of individuals involved in the battle.

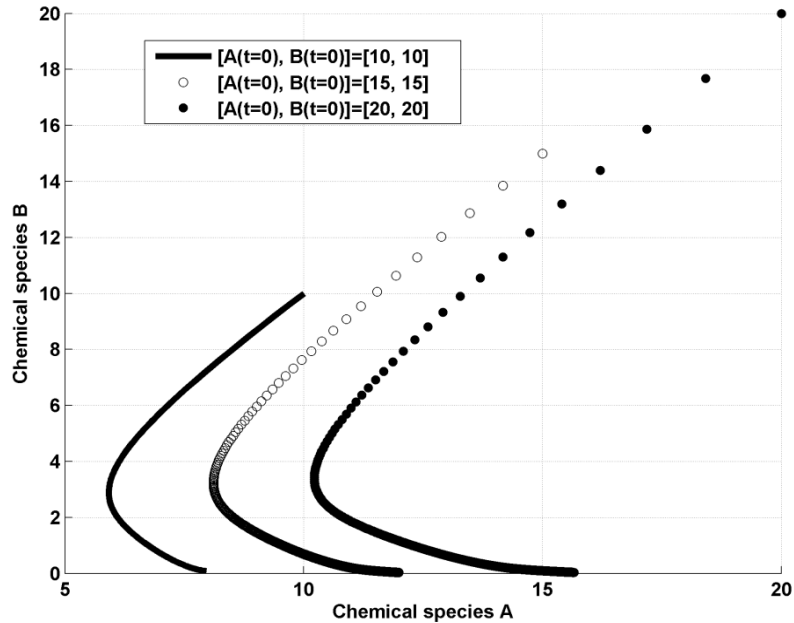

Fig. 2. Phase plane considering the chemical species A and B, obtained using different initial conditions.

Model parameters were estimated from one of the experiments where mortality of the two species occurred, using the Simplex Flexible Algorithm (SFA, [36]). This optimization algorithm find the best set of parameters $\mathbf{P} = [k_1, \ldots, k_i, \ldots, k_{15}]$ which minimize error function $F(\mathbf{P})$,

$$F(\mathbf{P}) = \sum_{j=1}^{5} \frac{1}{N} \sum_{n=1}^{N} \left( x_{j,n}^{\text{exp}} - x_{j,n}^{\text{mod}}(\mathbf{P}) \right)^2, \qquad (16)$$

where $x^{exp}$ and $x^{mod}$ indicate the experimental and the values obtained by the model, respectively, for each species $j=1, 2, \ldots, 5$ and for each population value $n=1, 2, \ldots, N$. Table 1 shows the estimated values of the parameters.

In Figure 3 and 4 we show observed changes in total population of species *A* and *B* and the corresponding values from the deterministic model with the optimized parameters, considering that the simulation time is 4620 seconds as in the reference experiment (for more details about population trend of each chemical species, see [39]).

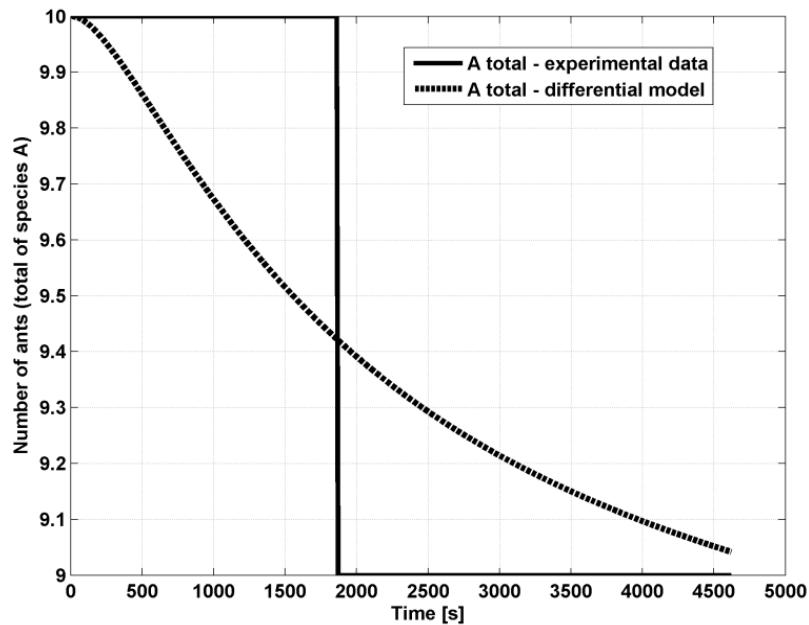

Fig. 3. Comparison between observed (continuous line) number of A individuals and predicted (dotted line) using the deterministic model with optimized parameters.

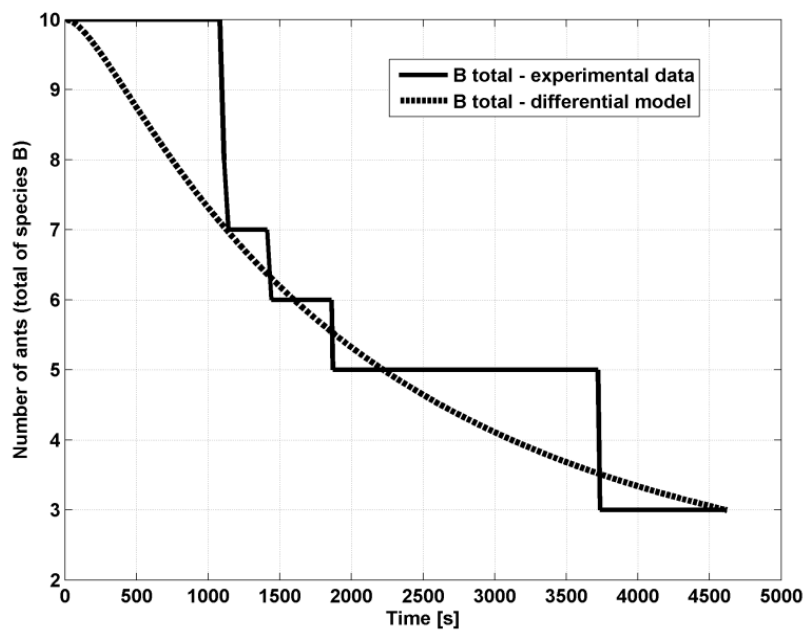

Fig. 4. Comparison between observed (continuous line) number of B individuals and predicted (dotted line) using the deterministic model with optimized parameters.

| $k_i$ | Optimized parameters by parametric identification of DE |
|---|---|
| $k_1$ | 0.0002438 |
| $k_2$ | 0.0006932 |
| $k_3$ | 7.7388e-05 |
| $k_4$ | 0.001211 |
| $k_5$ | 6.94257e-06 |
| $k_6$ | 4.75385e-06 |
| $k_7$ | 4.50353e-05 |
| $k_8$ | 3.40730e-05 |
| $k_9$ | 6.38355e-06 |
| $k_{10}$ | 1.05249e-05 |
| $k_{11}$ | 3.13041e-05 |
| $k_{12}$ | 0.00040597 |
| $k_{13}$ | 0.0009040 |
| $k_{14}$ | 1.22548e-06 |
| $k_{15}$ | 6.53508e-06 |

Tab. 1. Values of the best reaction constants $k_i$ obtained by means of the optimization algorithm SFA.

## 2.3 Standard Gillespie algorithm

Experimental data (see Figure 1) shows large fluctuations in the number of individuals and groups. We also observe wide fluctuations in the times at which the reactions occur. The deterministic model of Eqs. (11-15) cannot reproduce this behavior. The estimation of fluctuation is also important for establishing the minimum number of experimental observations needed to validate the parameters.

For this purpose we exploited the Gillespie's direct method [38], which is an algorithm that generates a statistically correct trajectory, given a set of transition probabilities. This method explicitly simulates each reaction, giving a stochastic formulation of the kinetics based on the theory of random encounters. Considering a time interval $\Delta t$, the reaction probability function is

$$\begin{cases} P(\Delta t, i) = P_1(\Delta t) \cdot P_2(i) \\ P_1(\Delta t) = F_0 \cdot \exp(-F_0 \cdot \Delta t) \\ P_2(i) = \dfrac{F_i}{F_0} \\ F_0 = \sum_{i=1}^{15} F_i \end{cases} \qquad (17)$$

where $F_i$ is the propensity function (see Eq. (20)).

The reaction probability density function is separable in two parts, an exponential distribution of time reactions $P_1$ and the normalized propensity function of reactions $P_2$. The Gillespie algorithm can be implemented in two steps by choosing the time interval $\Delta t$ and the identity $i$ of the reaction respectively as

$$\Delta t = \frac{1}{F_0} \cdot \ln\left(\frac{1}{r_1}\right), \tag{18}$$

and

$$\sum_{1}^{i^*-1} F_i < r_2 \cdot F_0 \leq \sum_{1}^{i^*} F_{i^*}, \tag{19}$$

where $i^*$ is the integer for which Eq. (19) is valid and $r_1$ and $r_2$ are uniform random numbers between 0 and 1. The propensity vector of reactions is

$$\begin{bmatrix} k_1 \cdot x_1 \cdot x_2 \\ k_2 \cdot x_3 \\ k_3 \cdot x_3 \\ k_4 \cdot x_3 \\ k_5 \cdot x_2 \cdot x_3 \\ k_6 \cdot x_4 \\ k_7 \cdot x_4 \\ k_8 \cdot x_4 \\ k_9 \cdot x_4 \\ k_{10} \cdot x_1 \cdot x_2 \cdot (x_2 - 1)/2 \\ k_{11} \cdot x_2 \cdot x_4 \\ k_{12} \cdot x_5 \\ k_{13} \cdot x_5 \\ k_{14} \cdot x_5 \\ k_{15} \cdot x_3 \cdot x_2 \cdot (x_2 - 1)/2 \end{bmatrix}, \tag{20}$$

where the factors express the number of different ways in which individuals can participate to the reaction.

In Figure 5 we show a sample of 100 stochastic trajectories, generated by means of the Gillespie algorithm, together with the mean-field prediction of the deterministic model, for the total number of species *A* and *B*. Figure 6 shows the frequency distributions of surviving A and B individuals obtained from 1000 simulated trajectories. The probability that 10 individuals of species *A* survive, is about of 45%, while for 9 individuals is about 35%. For the species *B*, the maximum probability of survival is about 30% corresponding to two individuals. Note that for the deterministic model we expect 9 survivors for the species *A* and 3 for species *B*, while the maximum survival probability (stochastic model) corresponds to 10 and 2 for species *A* and *B* respectively, due to finite size effects.

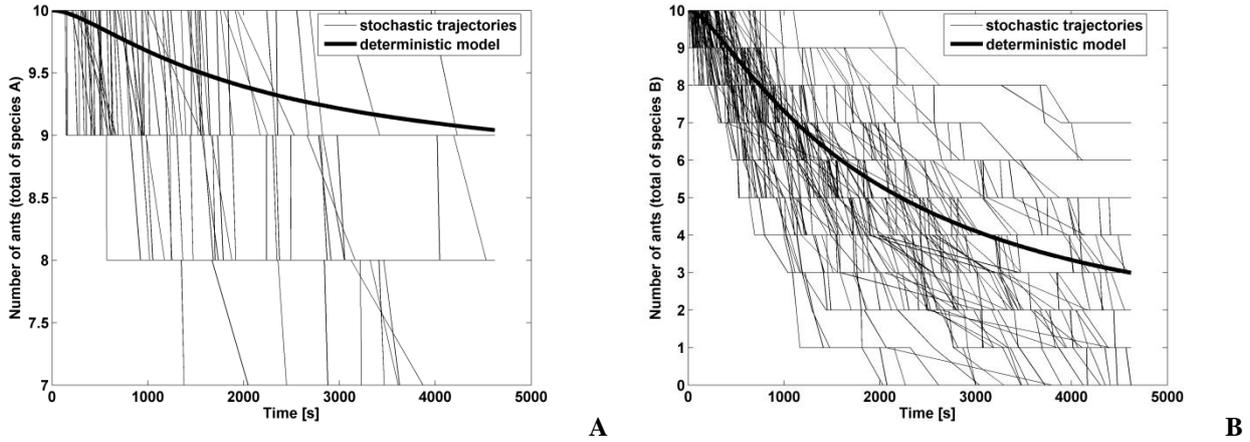

Fig. 5. Comparison between 100 stochastic trajectories and values predicted by the deterministic model. (A) Species *A*. (B) Species *B*.

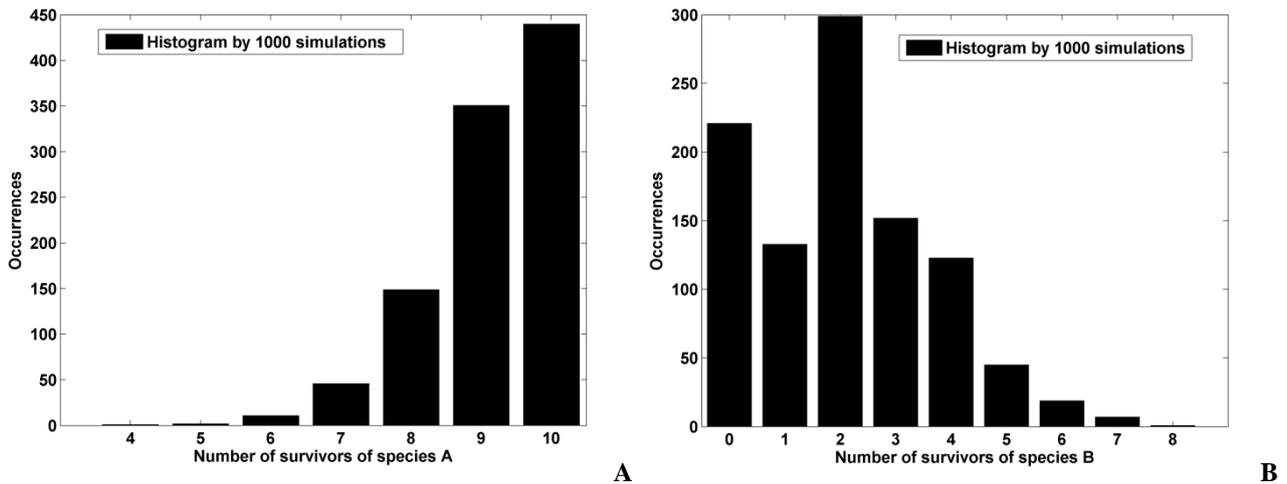

Fig. 6. Occurrence of survivors after 4620 seconds considering 1000 stochastic trajectories. (A) For the species *A*. (B) For the species *B*.

2.4 The diffusion-limited Gillespie algorithm

In order to verify the influence of the spatial dimension on ant battle dynamics we implemented a spatial agent-based model. We considered a disk (corresponding to a 10 cm Petri dish) where the ants of the two species move randomly. The idea is to replace the choice of random reaction times given by Eq. (18) with the collision probability of a random walk. Since it is difficult to estimate this probability in a continuous space represented by a disk (due to the presence of borders) we adopted a discrete lattice and we partitioned the disk into 1cm-wide square cells (see Figure 7). The chemical species *A* and *B* can diffuse to neighboring cells, while the fighting groups *AB*, *ABB* and *ABBB* remain in the cell where they formed, as observed in the experiments.

Ants can move from one cell to a neighboring or next-to-neighboring one with probability 1/8. This transition probability fixes the time scale of the system. As observed in experiments, "free" ants move continuously, unless they meet an opponent (*i.e.*, when individuals of different chemical species occupy the same cell). In this case the Gillespie dynamics is taken into account to determine the reaction probability,

$$\begin{cases} P(i) = \dfrac{F_i}{F_0} \\ F_0 = \sum_{i=1}^{15} F_i \end{cases}, \qquad (21)$$

with the propensity function expressed by Eq. (20).

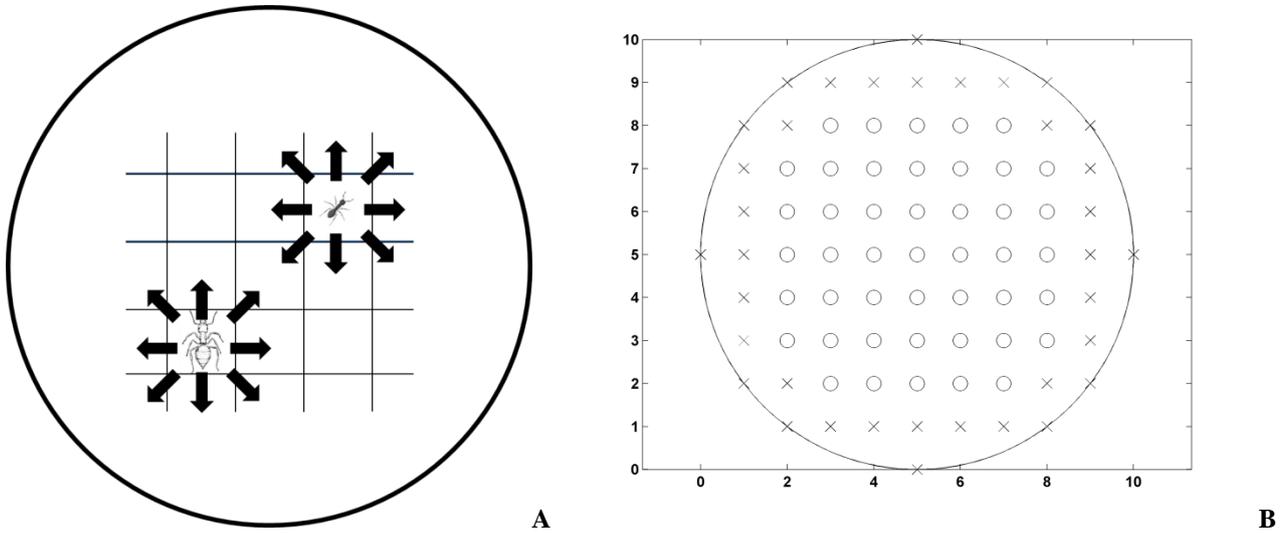

Fig. 7. (A) The virtual Petri dish, where the chemical species *A* and *B* move with a spatial step of 1 cm to the nearest and next-to-nearest neighbors according to 2D random walk. (B) The discrete lattice (81 cells) where the virtual ants move: the circles indicate the center of cells, while the crosses indicate the reflecting walls.

In terms of Markov chains we can express the probability to find a random walker in a given cell of the lattice, depending on its starting position, using the transition matrix **T** (Fig. (8)):

$$P_{i,t} = \mathbf{T}^t \cdot P_{i,0}, \qquad (22)$$

where $P_{i,t}=[p_{i,1}, p_{i,2},\ldots, p_{i,k},\ldots, p_{i,81}]$ represents the probability that a random walker *i* is located in one of the 81 cells of the lattice at time *t* giving its starting position $P_{i,0}$ at time 0 (see Figure 7B), depending on starting position $P_{i,0}$ in the time *t*. For example, if we start from position 40 we assume $p_{i,40}=1$ and $p_{i,k}=0$ for each $k\neq 40$. The transition matrix is implemented considering reflecting walls as indicated in Figure 7B.

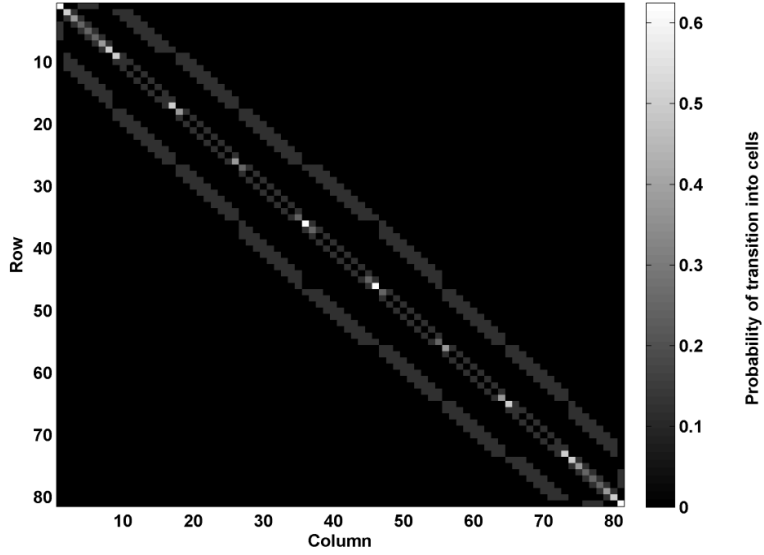

Fig. 8. Representation of the symmetric transition matrix **T**.

It is possible to express the probability $P_{ij,t}$ of the first intersection of two random walkers $(i, j)$, *i.e.*, two ants of different species, at time $t$ as the probability that the two trajectories have not met at previous times $(1, 2,...,t-1)$,

$$P_{ij,t} = \left[\prod_{n=1}^{t-1}(1 - P_{i,n} \cdot P_{j,n})\right] \cdot P_{i,t} \cdot P_{j,t}. \tag{23}$$

Assuming the independence of events, we can write the probability $I_t$ of the first intersection, given the number of random walkers, *i.e.*, the number of ants of the two species, as

$$I_t = \sum_{ij}\left\{P_{ij,t} \cdot \prod_{i,J \neq j}\left[1 - P_{i,t} \cdot P_{J,t}\prod_{n=1}^{t-1}(1 - P_{i,n} \cdot P_{J,n})\right] \cdot \prod_{I \neq i,j}\left[1 - P_{I,t} \cdot P_{j,t}\prod_{n=1}^{t-1}(1 - P_{I,n} \cdot P_{j,n})\right]\right\}. \tag{24}$$

Equation (24) represents the sum of the probabilities of pair intersections, considering for each pair $(i, j)$ of ants that each $i$ has not met any j $(J \neq j)$ and that each $j$ has not met any I $(I \neq i)$ before. We can include also the trajectories of ants that detach from a fighting group (for example considering the inverse reaction of Eq. (1)). Equation (24) determines the time in which is possible to have a reaction and replaces the exponential probability $P_1(\Delta t)$ of the classical Gillespie algorithm (see Eq. (17)). In Fig. 9 we show the transition probability of the first intersection for each cell of the virtual disk. These probabilities have a maximum between 2 and 11 time steps, depending on the initial position of ants.

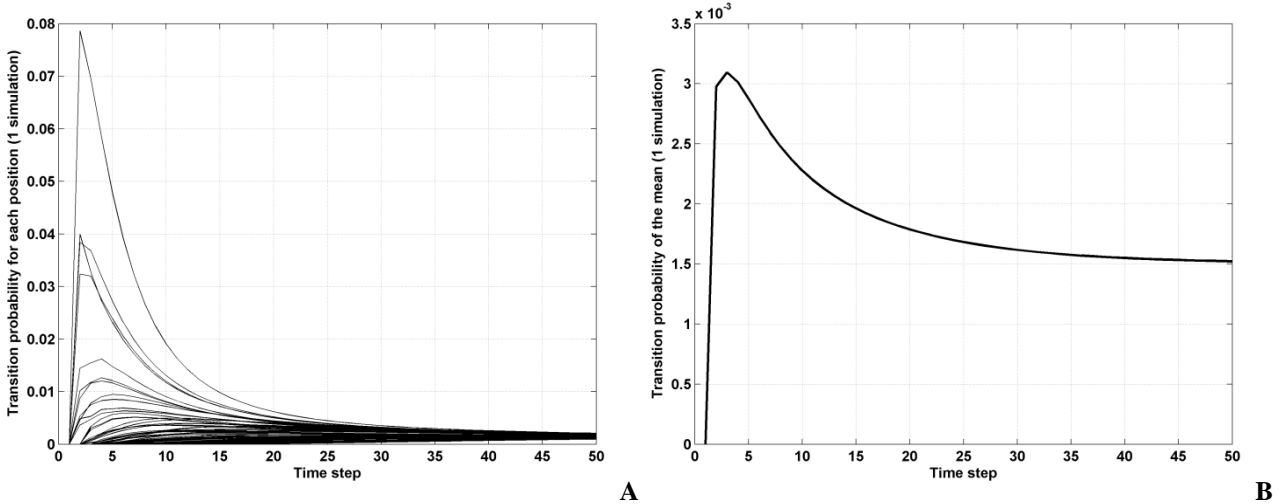

Fig. 9. (A) The transition probability of the first intersection over time for each position of virtual disk, calculated for one simulation. (B) The mean of transition probability of the first intersection over the time for one simulation.

In Figure 10 we show a simulation of the spatial model together with the position of the chemical species in the simulated Petri dish. In Figure 11 we report a comparison between 100 stochastic simulations, generated by means of Gillespie algorithm, and the numeric solution of DE for the total number of species *A* and *B*. As in the previous case, we can estimate the survival probability with a simple Monte Carlo simulation, considering 1000 trajectories (Figure 12). In this case the survival of 10 individuals of species *A*, is about 55%, while for 9 individuals is about 35%. For the species *B*, the maximum probability of survival is about 24% corresponding again to two individuals. The differences between the two statistics are due to stochastic fluctuations, to different probabilities of the corresponding models and to finite size effects.

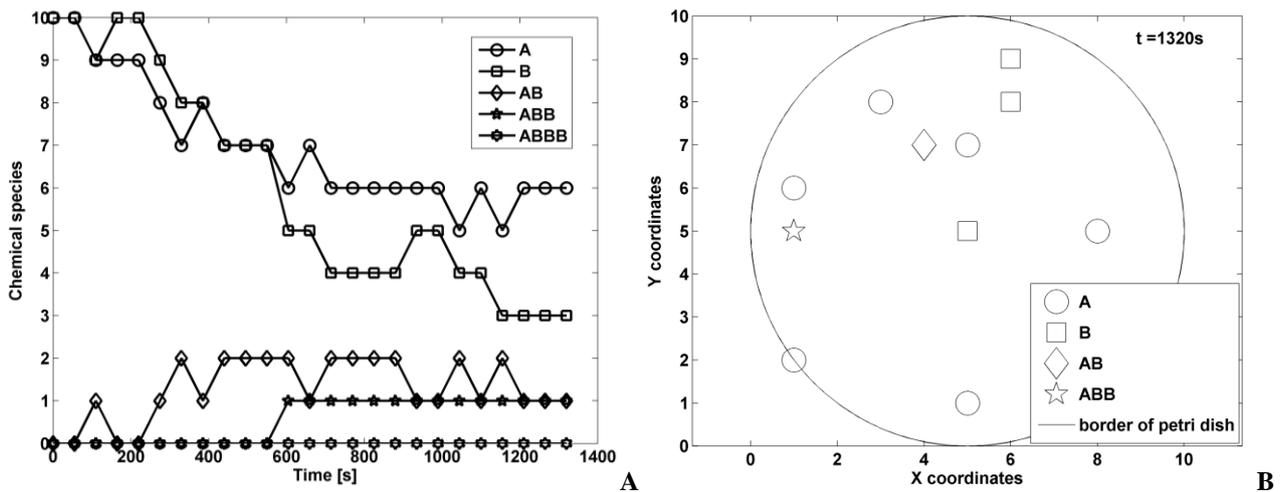

Fig. 10. A simulation of the Gillespie-based agent model. (A) The population trend over the time up to time t = 1320s. (B) The positions of the chemical species in a virtual Petri dish at time t = 1320s.

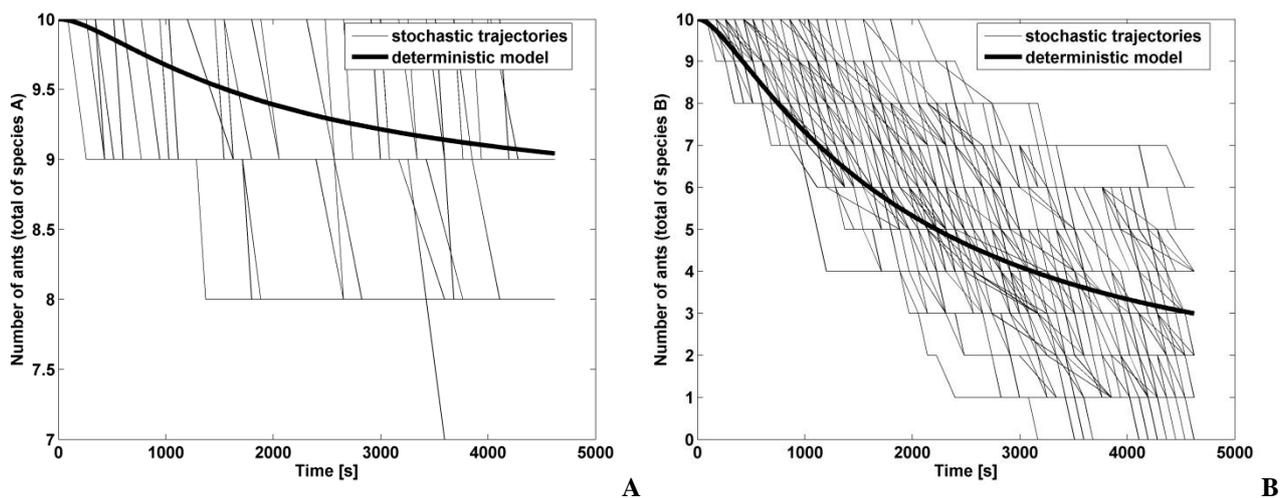

Fig. 11. Comparison between predictions of the deterministic model and 100 stochastic trajectories generated by means of Gillespie-based agent model. (A) Species *A*. (B) Species *B*.

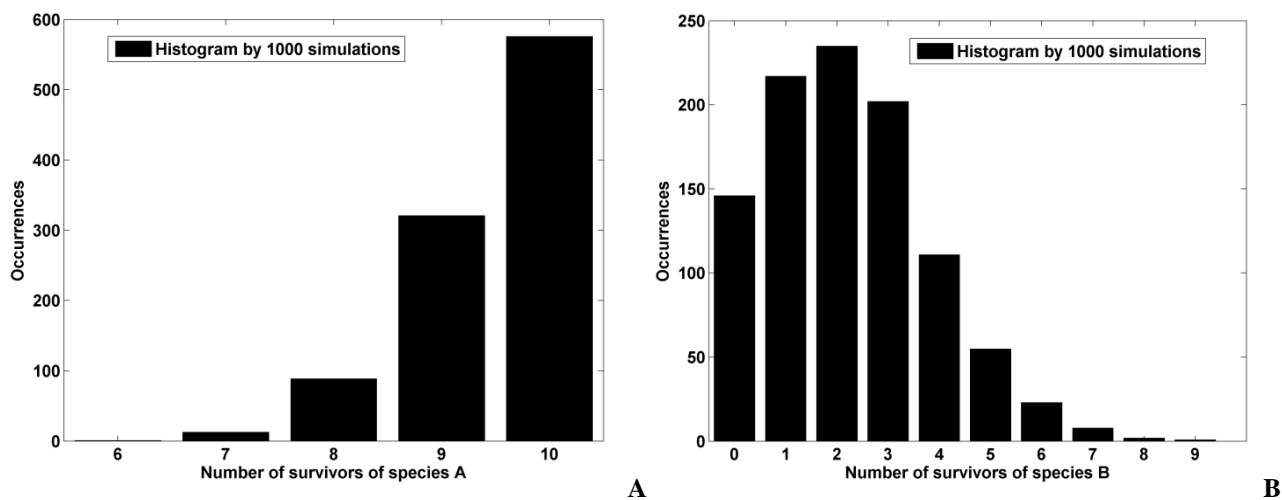

Fig. 12. Occurrence of number of survivors considering 1000 stochastic trajectories generated by means of Gillespie-based agent model. (A) Species *A*. (B) Species *B*.

## 3. Discussion

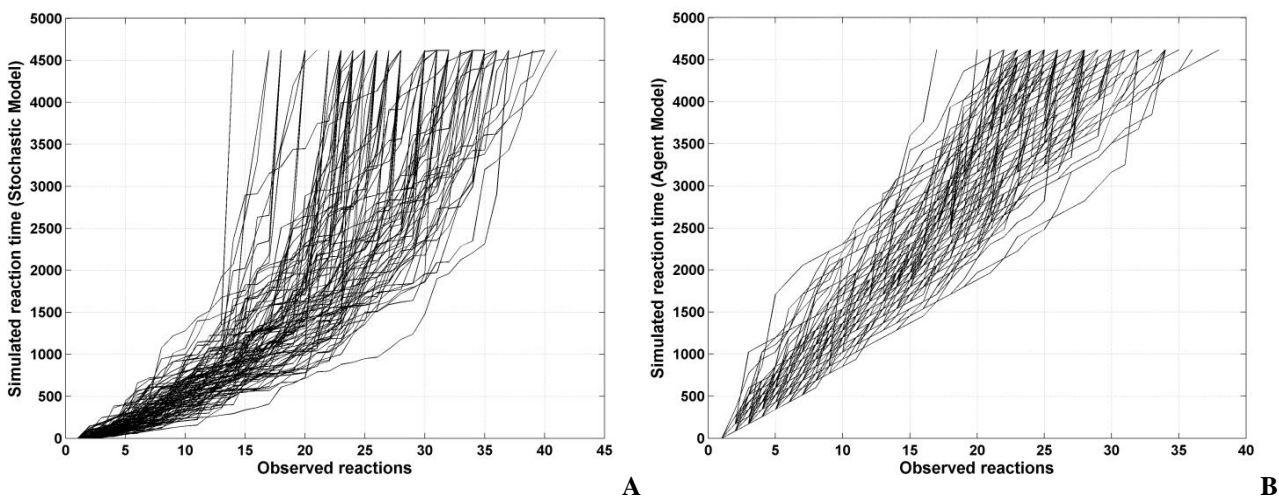

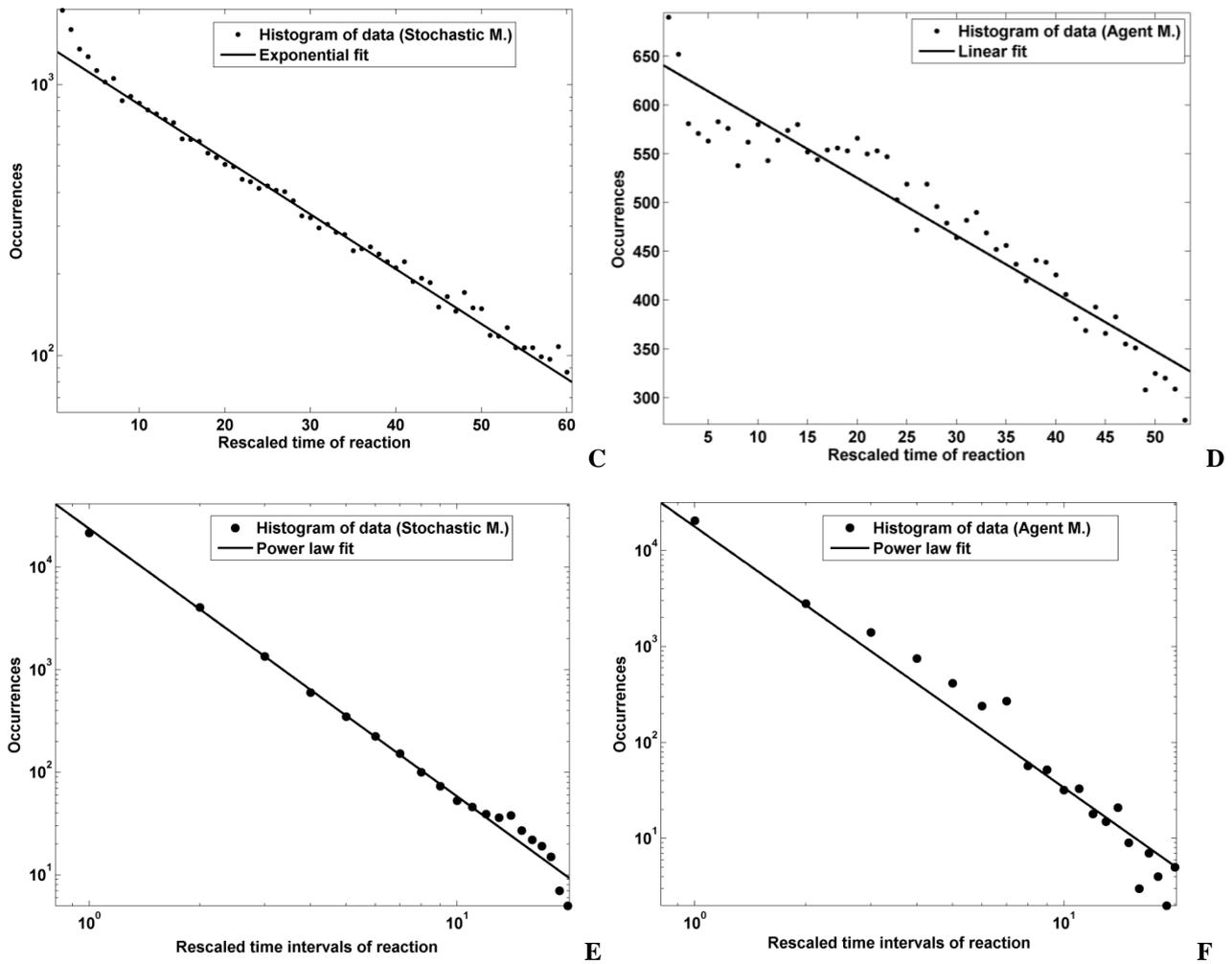

Fig. 13. Comparison between the time-dependent stochastic model and the spatial Gillespie-based agent model from the statistical point of view. (A) Time of reaction (100 simulations) vs. the observed reactions for the time-dependent stochastic model. (B) Time of reaction (100 simulations) vs. the observed reactions for the spatial Gillespie-based agent model. (C) The distribution of the time reaction (1000 simulations) for the time-dependent stochastic model. (D) The distribution of the time reaction (1000 simulations) for the spatial Gillespie-based agent model. (E) The distribution of the time reaction intervals (1000 simulations) for the time-dependent stochastic model. (F) The distribution of the time reaction intervals (1000 simulations) for the spatial Gillespie-based agent model.

Considering the simulations, both the standard and the diffusion-limited Gillespie models are suitable to describe the experiments. The main difference between the two models, as mentioned above, is the probability of the intersection of the random walkers that determines the time of a possible reaction (Figure 13B), while in the standard Gillespie model the time of reaction is chosen considering an exponential distribution (Figure 13A). We performed 1000 simulations to compare the statistics of the reaction times of the two models. The standard Gillespie model exhibits, as expected, an exponential distribution (R-square = 0.996) of the reaction times (Figure 13C), while the diffusion-limited one exhibits, with a sufficient approximation, a linear distribution (R-square = 0.918). This difference is evident in Figure 13A and 13B. The standard Gillespie model promotes initially a faster dynamics than the diffusion-limited one, based on random walks, as can be seen comparing Figure 5 (initially fast dynamics) with Figure 11 (regular dynamics due to probability of random walk intersection). The distribution of the time increments is always a power law distribution. For the standard Gillespie model we achieve an exponent γ = -2.611 with a R-square = 1, while for the diffusion-limited one an exponent γ = -2.719 with a R-square = 0.9998.

## 4. Conclusions

We performed laboratory experiments to model ant battle dynamics and the fighting strategies of two ant species, *Lasius neglectus* and *Lasius paralienus*. We built two models, a simple time-dependent model and a spatial one, both based on a stochastic approach. The first one is based on the Gillespie algorithm that allows to choose the reactions (Eqs. (1-10)) and the time in which the reaction occurs. The second model (the spatial one), is based on the Gillespie algorithm only for the choice of the reaction, while the reaction time depend on the probability of the intersection between the random walks. Analyzing the simulations, we can conclude that both stochastic models are suitable to describe the ant battle dynamics, and that more experiments are necessary in order to appropriately validate the models.

The comparison between the statistical analysis of two models shows that the introduction of space in the dynamics affects the trend of the reaction times (see Figure 13A and 13B). One of the most important differences is that, initially, in the standard Gillespie model there is a faster reaction dynamics, compared to the diffusion-limited one, due to the effects of the exponential reaction time of the Gillespie algorithm (Figure 13A and 13B). Secondly, the effect of modeling ant motion as random walks leads to a linear distributions of reaction times (see Figure 13D) and thus almost-linear trajectories (Figure 13B).

The main results of our approach is that, by means of the Monte Carlo simulations, we can estimate the probability of survival of the two considered species in the battle.

We are preparing and analyzing several new experiments in order to verify and validate our methodology. The results of the analysis and the comparisons with the models will be presented elsewhere.